\documentclass[12pt]{iopart}
\usepackage{amsfonts}
\begin{document}
\jl{1}
\title{UV divergence-free QFT on noncommutative plane}
\author{Anais Smailagic\dag{}\footnote[1]{e-mail
        address: \texttt{anais@ictp.trieste.it}},
        Euro Spallucci\ddag{}\footnote[5]{e-mail
        address: \texttt{spallucci@trieste.infn.it}}
        }
        \address{\dag{}Sezione INFN di Trieste,\\
         Strada Costiera 11, 34014 Trieste,\\
         Italy}
        \address{\ddag{}Department of Theoretical Physics,\\
         University of Trieste, Strada Costiera 11, 34014 Trieste,\\
         Italy}

\date{\today}

\begin{abstract}
We formulate Noncommutative Qauntum Field Theory in terms of fields defined 
as mean value over coherent states of the noncommutative plane. 
No $\ast$-product is needed in this formulation and noncommutativity is carried
by a modified Fourier transform of fields. As a result the theory is UV finite 
and the cutoff is provided by the noncommutative parameter $\theta$.
\end{abstract}

\pacs{11.10.Nx}
\maketitle

   Recent revival of interest in noncommutative theories has been triggered by
   the results in string theory. The early results in this subject \cite{witten}
   \cite{sw},\cite{schw}, \cite{nekra},
   have been followed by a vast number of papers dealing with the problem of 
   formulating a noncommutative quantum mechanics \cite{tutti} and field theory. 
   Though it is possible to formally define these models \cite{agm} it is
   hard to perform any calculation directly in terms of noncommutative
   variables. It has emerged that the most promising approach is to ``simulate''
   noncommutativity in the space of ordinary functions by the use of 
   $\ast$-product, as it has been attempted long time ago in ordinary quantum
   mechanics \cite{WWM}. We have recently shown that the same results can be
   obtained by suitable redefinition of noncommutative coordinates in terms
   of canonical ones \cite{noi}. In this case, the effect of noncommutativity
   manifests itself as an ``external, constant magnetic field''. \\ 
   On the other hand, quantum field theory has been so far formulated only by
   replacing ordinary commutative product of fields by a $\ast$-product in the 
   original Lagrangian \cite{agm},\cite{nekra}. This is due to the fact that  
   field theories are formulated in Lagrangian formalism, unlike quantum 
   mechanics which uses Hamiltonian formulation allowing to work with phase 
   space coordinates. This difference does not allow a straightforward 
   extension of nice and simple description available in quantum mechanics. 
   In fact, one has to find a self-consistent way of treating noncommutativity 
   of coordinates only.\\
   With the above consideration in mind, we would like to remark that
   a fundamental effect of noncommutativity is the change in nature
   of the coordinate space which becomes \textit{blurry} because of 
   the existence of a minimal length determined the noncommutative parameter 
   $\theta$. In momentum space description of quantum field theory one expects 
   a natural cutoff provided by $\theta$, thus rendering the theory UV finite. 
   Similar ideas, motivated by quantum gravity effects have been introduced
   in \cite{padma}.
   UV finiteness should be 
   the first test of a successfully formulated 
   Noncommutative Quantum Field Theory(NCQFT).
   Every  paper carries this basic expectation, but it 
   has not been achieved if one deals with $\ast$-product. The reason is that,
   in order to perform calculations, 
   one is forced to  expand in $\theta$ the $\ast$-product and to keep 
   only the first few terms. As it has been asserted in many papers, the
   divergences of Feynman diagrams are not cured by the $\theta$-parameter.
   Instead of UV finiteness one finds again UV divergences of the commutative 
   field theory. The only effect of noncommutativity in this approach
   is to generate new, non-planar, Feynman diagrams. Final result is UV/IR 
   mixing which is defying Renormalization Group expectations.\\
   In this letter we shall present a way to reformulate NCQFT which 
   incorporates $\theta$ as a natural cutoff both in propagators and vertices, 
   in such a way to render the theory UV finite.\\
   We shall describe a noncommutative scalar field theory on $2+1$ dimensional
   spacetime. We follow usual wisdom to keep time as a commutative
   coordinate in order to avoid problems with unitarity \cite{cdpt}, while
   space coordinates are noncommutative. We choose to work in a 
   plane being the simplest noncommutative geometry.\\
   The noncommutative plane is described by space coordinates satisfying
   commutation rules given by
   
   \begin{eqnarray}
  && \left[\, \mathbf{X}^i\ ,\mathbf{X}^j\,\right]= i\, \theta\, \epsilon^{ij}
   \qquad i\ , j= 1,2\label{comm}\\
   && \left[\, \mathbf{X}^i\ ,\mathbf{P}_j\,\right]= i\, \delta^i_j\\
   &&\left[\, \mathbf{P}_i\ ,\mathbf{P}_j\,\right]= 0
   \end{eqnarray}  
  
   where we have
   chosen  units $\hbar=c=1$. $\theta$ has dimensions of a length squared
   and  measures  the noncommutativity of coordinates. Conjugate
   momenta $\mathbf{P}_i$ are chosen to satisfy standard commutation rules.
   As a consequence of (\ref{comm}), the noncommutative plane  is
   divided into plaquettes of area $\theta$. One cannot speak of points anymore
   and the space becomes blurry. \\
   This immediately leads to the question: how to define a function (field)
   of space coordinates?\\
   In our view, the main point in formulating a NCQFT
   is to find a proper set of states which allow to define mean value of
   a function $F\left(\, \mathbf{X}^1\ ,\mathbf{X}^2 \,\right)$. 
   The difference with respect to commutative theory stems from
   the fact that $\mathbf{X}^1$ and $\mathbf{X}^2$   are operators having
   \textit{no common position eigenvectors} $\vert\, x^1\, x^2\, \rangle $
    due to (\ref{comm}). In order to look for a convenient set of states
    let us introduce a  set of operators defined as
    
    \begin{eqnarray}
   && \mathbf{Z}\equiv \frac{1}{\sqrt{ 2}}\left(\, \mathbf{X}^1 + i
   \mathbf{X}^2\,
   \right)\\
   && \mathbf{Z}^\dagger\equiv \frac{1}{\sqrt{ 2}}\left(\, \mathbf{X}^1 - i
   \mathbf{X}^2\, \right)
   \end{eqnarray}
    
    The new operators satisfy commutation relation
   
   \begin{equation}
   \left[\, \mathbf{Z}\ , \mathbf{Z}^\dagger\,\right]=  \theta\ ,
   \label{comm2}
   \end{equation}  
   
   One can recognize that the $\mathbf{Z}$   $\mathbf{Z}^\dagger$  operators
   satisfy the commutation relation of creation/annihilation operators, of
   ordinary quantum mechanics, with the formal substitution
   $\hbar\longrightarrow \theta$. Thus, the commutative limit $\theta \to 0$     
   corresponds to the classical limit, $\hbar\to 0$, of quantum mechanics.
   It is known, since the seminal work of Glauber in quantum optics
   \cite{glaub}, that there exist \textit{coherent states} which are
   eigenstates of annihilation operator. 
   The advantage of working with  operators  $\mathbf{Z}$ ,  
   $\mathbf{Z}^\dagger$, in place of $\mathbf{X}^1$ and $\mathbf{X}^2$,
    is that there exists eigenstates satisfying 
  
   \begin{eqnarray}
   &&\mathbf{Z}\, \vert\, Z\, \rangle = z\, \vert\, Z\, \rangle \label{z1}\\
   &&\langle\, Z\,\vert\, \mathbf{Z}^\dagger = \langle\,
   Z\,\vert\, \bar z \label{z2}
   \end{eqnarray}
   
   having complex eigenvalues $z$. The explicit form
   of the normalized $\vert\, Z\, \rangle $ states is

   \begin{equation}
    \vert\, Z\, \rangle\equiv \exp\left(\, -\frac{z\,\bar z  }{2\theta}\,\right)
   \exp\left(\, -\frac{z }{\theta}\, \mathbf{Z}^\dagger\,\right)\vert\, 0\, 
   \rangle
   \end{equation}
   
   where, the vacuum state $\vert\, 0\,  \rangle $ is annihilated by 
   $\mathbf{Z}$.\\
  The $\vert\,\mathbf{Z}  \,  \rangle $ states are the coherent states 
  of the noncommutative plane and satisfy the completeness relation
  
   \begin{equation}
   \frac{1}{\pi\theta}\int dz\, d\bar z\, \vert\, z\, \rangle \langle\, z\,
    \vert = 1
   \end{equation}
   
   Coherent states allow to associate to any operator $F\left(\, \mathbf{X}^1\ ,
   \mathbf{X}^2 \,\right)$ an ordinary function $F(z)$ as
  
   \begin{equation}
   F(z)\equiv  \langle\, z\, \vert\,  
   F\left(\, \mathbf{X}^1\ , \mathbf{X}^2 \right)\,
   \vert\, z\,  \rangle \label{media}
   \end{equation}
   
   with the use of the mean value of $\mathbf{X}^1$,$\mathbf{X}^2$ given by
   
    \begin{eqnarray}
   &&\langle\, z\, \vert\,   \mathbf{X}^1\, \vert\, z\,  \rangle=
   {\sqrt 2}\Re z\label{x1}\\
   &&\langle\, z\, \vert\,   \mathbf{X}^2\, \vert\, z\,  \rangle=
   {\sqrt 2}\Im z\label{x2}
   \end{eqnarray}
   
   Above definitions open  road to a definition of the quantum
   fields on the noncommutative plane. Let us first define the noncommutative
   version of the Fourier transform 
   \begin{equation}
   F\left(\, z\, \right)=\int \frac{d^2p}{2\pi}\, f\left(\, p
   \,\right)\,  \langle\, z\, \vert  \exp\left(\, i p_j \mathbf{X}^j \, \right)
   \vert\, z\,  \rangle
   \end{equation}
   
  With the help of (\ref{x1}), (\ref{x2}) the mean value of noncommutative 
  plane wave can be rewritten as
  
  \begin{eqnarray}
  \fl \langle\, z\, \vert \, \exp\left(\, i p_j \, \mathbf{X}^j \, \right) \,
   \vert \, z \, \rangle &&=
   \langle\, z\, \vert \, \exp\left(\, i p_+ \, \mathbf{Z}^\dagger \, \right) 
   \,\exp\left(\,  p_- \, \mathbf{Z} \, \right) \,\exp\left(\, \frac{p_-p_+}{2} \,
   \left[\, \mathbf{Z}^\dagger\ , \mathbf{Z}\,\right] \, \right)
   \vert \, z \, \rangle \nonumber\\
   &&\label{onda}
   \end{eqnarray}
   
   where, $p_\pm\equiv \left(\, p_1\pm ip_2\,\right)/\sqrt 2$. 
   We  have used the Hausdorff decomposition of the exponent due the
   noncommutativity of the coordinates which introduces additional
   factor in the definition of the plane wave on the noncommutative plane.
   This is the crucial point of our method, i.e. the noncommutativity is
   seen as a modified Fourier transform of ordinary functions, given by
   
   \begin{equation}
  \fl F(z)=\int \frac{d^2p}{2\pi}\, f\left(\, p\,\right)\,
   \exp\left[\, - \frac{\theta }{ 4}\, \left(\, p_1^2 + p_2^2 \,
   \right)\,\right] 
   \exp\left[\, +i \frac{ p_1}{\sqrt 2 }\left(\, z + \bar z\,\right)  
   + \frac{ p_2}{\sqrt 2 }\left(\, z - \bar z\,\right)  
   \,\right] \label{fourier}
   \end{equation}
   
   The above result shows that noncommutativity  produces a gaussian dumping 
   factor. To emphasize the difference between commutative and 
   noncommutative case, let us choose $f\left(\, p\,\right)=
   \mathrm{const.}$ corresponding to the maximum spread in momentum.
   The Fourier transform gives 
   
   \begin{equation}
   F(z)=\frac{4\pi}{\theta}\exp\left[\, -\frac{4}{\theta}\, z\bar z\,\right]
   \end{equation}
   Thus, we find a gaussian distribution. The
   reason is that the gaussian ``remembers'' the noncommutativity of the
   space. Even if the momentum has maximal
   spread, the uncertainty of the coordinates can shrink only to a minimal
   width proportional to $\sqrt\theta$, indicating blurriness of space.
   In the commutative limit $\theta\to 0$ one recovers the usual
   Dirac delta function.
   As a result of the above discussion, one can assert that the noncommutativity
   can be introduced in the Fourier transform by replacing ordinary plane waves
   by gaussian wavepackets. \textit{A posteriori} this conclusion sounds quite
   natural.\\   
   Now we are ready to define a quantum field on a noncommutative plane.
   A scalar field of mass $m$ will be described  through the expansion
   
   \begin{equation}
   \phi\left(\, t\ ,z\,\right)=
   \sum_{E , p}\left[\, \mathbf{a}^\dagger_p\exp\left(-i\,E\, t\,\right) 
   \langle\, z\, \vert \,
   \exp\left(\, i p_j \, \mathbf{X}^j \, \right) \, \vert \, z \, \rangle 
   +\mathrm{h.c.}\,\right]
   \end{equation}
   
   where, $\mathbf{a}^\dagger_p$,$\mathbf{a}_p$ are usual creation/annihilation
   operators acting on  Fock states with definite energy and momentum. They are
   the same as in the commutative case since the momenta commute among
   themselves.\\
   Armed with the above definitions, let us  compute the noncommutative 
   version of the Feynman propagator,  which is
   
   \begin{eqnarray}
  \fl G\left(\, t_1 - t_2\ , z_1-z_2\,\right)&&= 
  \langle\, \vec p= \vec 0\, \vert T\left[\, \phi\left(\,t_1\ , z_1\,\right)
   \phi\left(\,t_2\ , z_2\,\right)\,\right]\vert\, \vec p=\vec 0 \,\rangle
   \nonumber\\
   &&=\int \frac{dE}{\sqrt{2\pi}}\exp\left[\, -i\,E\,\left(\,
   t_1-t_2\,\right)\,\right]
   \int \frac{d^2p }{2\pi } G\left(\, E\ , \vec p^{\, 2}\,\right)
   \times\nonumber\\ 
   &&\exp\left[\,i \frac{ p_1}{\sqrt 2 }\left(\, z_1 + \bar z_1
   -z_2 - \bar z_2 \,\right)  
   + \frac{ p_2}{\sqrt 2 }\left(\, z_1 - \bar z_1  - z_2 +\bar z_2 \,\right)  
   \,\right] \label{dg}
   \end{eqnarray}
   
   where the momentum space propagator is 
    
    \begin{equation}
    G\left(\, E\ , \vec p^{\, 2}\,\right)\equiv \frac{1}{-E^2+ \vec p^{\, 2} 
    + m^2  }
   \exp\left(\, - \frac{\theta }{ 2} \vec p ^{\, 2} \,\right)
   \end{equation}
   The above result nicely displays the expected UV cutoff arising
   from the noncommutativity of the coordinates. Thus, in our approach
   the effect of noncommutativity that everyone expects
   is achieved with the help of coherent states. We would like to remark that
   the modification of the   Fourier transform follows from the definition
   of the mean value over coherent states and is not an \textit{ad hoc} 
   construction.\\   
   Having constructed a dumped Feynman propagator (\ref{dg}), we want to find
    the corresponding Green function equation. We find it to be:

    \begin{eqnarray}
  && \left[\,-\partial^2_t+ \partial_{\, z_1}\, \partial_{\, \bar z_1} + m^2\,
   \right]\, G\left(\,t_1-t_2\ , z_1 - z_2\,\right)=
   \delta\left(\, t_1-t_2\,\right)\,\times\nonumber\\
   &&\frac{2\pi}{\theta} \exp\left[\,-\frac{1}{4\theta}\left(\, z_1 + \bar z_1
   -z_2 - \bar z_2 \,\right)^2 +
   \frac{1}{4\theta}\left(\, z_1 - \bar z_1
   -z_2 + \bar z_2 \,\right)^2
   \,\right]
     \end{eqnarray}
   
   Again, we find a natural extension in the  noncommutative plane
   i.e. Dirac delta function of coordinates is replaced by a gaussian function.
   The commutative result is recovered as $\theta\to 0$.\\
   Based on  previous results, we define the 
   Lagrangian of the noncommutative scalar field as
   
   \begin{equation}
   L=\frac{1}{2}\left[\, \left(\partial_t\phi\,\right)^2 -\partial_{\bar z}\phi
   \partial_{ z}\phi +m^2\phi^2\,\right] - V\left(\, \phi\, \right)
   \label{lagr}
   \end{equation}
   
   It is important to point out that the product of fields in (\ref{lagr})
   is not a $\ast$-product, but ordinary product of functions. We do not
   need a  $\ast$-product since the noncommutativity is embedded in the
   Fourier transform of a single field (\ref{fourier})  rather than in the 
   product among fields.
   The Feynman rules following from (\ref{lagr}) are the standard ones, except
   that both vertices and propagators are endowed with their own gaussian
   dumping factors ensuring UV finiteness of the theory. Localization of
   noncommutative effects within the Fourier transform of single field
   avoids unwanted cancellation among gaussian factors which
   has  taken place when working with $\ast$-products \cite{presn}.\\
   
   Note added in proofs.\\
   While this paper was in the process of refereeing, another
   paper of ours, written after this one, was already published in \cite{ll}.
   We discuss in \cite{ll} the formulation of the Feynman path integral using 
   the coherent state formalism introduced here.


\Bibliography{99}
\bibitem{witten}  E. Witten
Nucl. Phys. B\textbf{460} 335  (1996) 
\bibitem{sw} N.Seiberg,  E. Witten
JHEP \textbf{9909}  032 (1999)
\bibitem{schw} A. Konechny, A. Schwarz
  Phys.Rept. \textbf{360} (2002) 353-465
 \bibitem{nekra}M. R. Douglas, N. A. Nekrasov
 Rev.Mod.Phys. \textbf{73} (2001) 977
\bibitem{padma} T.Padmanabhan, Class. Quant. Grav. \textbf{4} (1987), L107;
T. Padmanabhan,  Phys. Rev. Lett.\textbf{78} (1997), 1854 
\bibitem{tutti}
R. P. Malik, A. K. Mishra, G. Rajasekaran
Int. J. Mod. Phys. \textbf{A13}4759 (1998);
V.P. Nair Phys. Lett. B \textbf{505} 249 (2001); 
V.P. Nair, A.P. Polychronakos 
Phys. Lett. B \textbf{505}  267 (2001);
S.Bellucci, A. Nersessian, C.Sochichiu 
 Phys.Lett. B522 (2001) 345; 
R.Banerjee, S.Kumar    
Phys. Rev.D\textbf{60}   085005  (1999)
\bibitem{agm}L.Alvarez-Gaume, S. R. Wadia
 Phys. Lett. B \textbf{501} 319 (2001);
 L. Alvarez-Gaume, J.L.F. Barbon 
 Int. J. Mod. Phys.\textbf{A16} 1123  (2001) 
\bibitem{WWM} 
H.Weyl Z.Phys. \textbf{46} 1 (1927);
E.P. Wigner  Phys. Rev.\textbf{40} 749  (1932);
G.E. Moyal Proc. Camb. Phyl. Soc. \textbf{45} 99 (1949)
\bibitem{noi}
 A.Smailagic, E.Spallucci
 J.Phys. \textbf{A35} (2002) L363;
 A.Smailagic, E.Spallucci
Phys. Rev.\textbf{D65} 107701 (2001)
\bibitem{cdpt} M. Chaichian, A. Demichev, P. Presnajder, A. Tureanu
Eur. Phys. J. \textbf{C20} (2001) 767
\bibitem{glaub} R.J. Glauber 
Phys. Rev.\textbf{131} 2766  (1963)
\bibitem{presn} M. Chaichian, A. Demichev, P. Presnajder
Nucl. Phys. B\textbf{567 } 360   (2000)
\bibitem{ll} A.Smailagic, E.Spallucci,
 J.Phys.\textbf{A36}  L467  (2003)
\end{thebibliography}
\end{document}